\documentclass[a4paper,12pt]{article}
\usepackage{a4wide}
\usepackage[bookmarks=false,colorlinks]{hyperref}
\usepackage[pdftex]{graphicx}
\usepackage{amsmath,amsfonts,amssymb,mathtools,bm}
\usepackage{rotating,amsmath,bm}
\usepackage{tabularx}
\usepackage[table, dvipsnames]{xcolor}
\usepackage{multicol,multirow}
\usepackage{latexsym}
\usepackage{relsize}
\usepackage[center]{subfigure}
\usepackage{float}
\usepackage{caption}
\usepackage{slashed}
\usepackage{lineno}
\usepackage{cite,mcite}
\usepackage{appendix}
\usepackage{url}
\usepackage{placeins}
\hypersetup{urlcolor=blue, citecolor=blue}
\usepackage[english]{babel}
\usepackage{bbold}
\usepackage{wasysym}
\usepackage[normalem]{ulem}
\usepackage{tikz}
\usepackage{cite}
\usetikzlibrary{arrows}
\tikzstyle{int}=[draw, minimum size=3em]
\tikzstyle{init}=[pin edge={to-, thick, black}]

\graphicspath{{wtf_01_gate0/}, {wtf_01_gate99/}, {wtf-uliana/gate-0/T=-35/}, {wtf-uliana/gate-0/T=-45/}, {wtf-uliana/gate-0/T=-55/}, {wtf-uliana/gate-3ff/}, {wtf-uliana/gate-0/bias-dependence/vb=70.28/}, {wtf-uliana/gate-0/bias-dependence/vb=70.37/}, {wtf-uliana/gate-0/bias-dependence/vb=70.44/}, {wtf-uliana/gate-0/comparator/vth=1/}, {wtf-uliana/gate-0/comparator/vth=1.2/}, {wtf-uliana/gate-0/comparator/vth=1.35/}}
\DeclareGraphicsExtensions{.pdf,.png,.jpg}
\usepackage{wrapfig} % Обтекание рисунков и таблиц текстом

\date{}
\normalsize

\begin{document}

%\linenumbers

\title{\bf The avalanche delay effect in sine-gated single-photon detector based on InGaAs/InP SPADs}

\maketitle

{\centering \sc
Alexandr Filyaev${}^{1,2,3}$, Anton Losev$^{1,2,4}$, Vladimir Zavodilenko$^{1,2,3}$, Igor Pavlov$^{1,2}$
\par}

\vspace{5mm}
{\centering \sl\small \noindent
${}^1$"QRate" LLC, St. Novaya, d. 100, Moscow region, Odintsovo, Skolkovo, 143026, Russia. \\
${}^2$National University of Science and Technology MISIS, Leninsky prospect, 4, Moscow, 119333, Russia. \\
${}^3$HSE University, Myasnitskaya ulitsa, 20, Moscow, 101000, Russia. \\
${}^4$National Research University of Electronic Technology MIET, Shokin Square, 1, Zelenograd, 124498, Russia. \\
\par}

\vspace{1cm}

\begin{abstract}
   A sine-gated single-photon detector (SPD) intended for use in a quantum key distribution (QKD) system is considered in this paper. An "avalanche delay" effect in the sine-gated SPD is revealed. This effect consists in the appearance of an avalanche triggered at the next gate after the photon arrival gate. It has been determined experimentally that the nature of this effect is not related to the known effects of afterpulsing or charge persistence. This effect negatively affects the overall error rate in the QKD system. The influence of the main detector control parameters, such as temperature, gate amplitude and comparator’s threshold voltage, on the avalanche delay effect was experimentally established.  
\end{abstract}

\thispagestyle{empty}

%\pagebreak
\newpage
\setcounter{page}{1}

%\linenumbers

\section{Introduction}

A single-photon detector (SPD) is a device capable of sensing single photons at a specific wavelength. Such a device has many applications \cite{yu2017fully, lee201672, zhang2016enhanced, al2019multimodal,kiktenko2018quantum, ceccarelli2017development}, but the most promising application is in quantum key distribution (QKD) \cite{huang2022dependency}. There are several types of devices that can be used as single photon detectors \cite{liang2016room}. The optimum device to create a miniaturised SPD and a compact QKD system as a whole is the InGaAs/InP based single-photon avalanche diode (SPAD). 

It is important to minimise the level of SPD false triggers, which entails an increased error rate for a QKD system with such a detector as part of it. One way to keep noise to a minimum is to set the control parameters of the detector correctly. 

One recently discovered effect is the avalanche delay effect, which causes false triggering of a sine-gated SPD in an adjacent gate. In this paper the influence of the detector control parameters on this negative effect is established experimentally.

\newpage
\section{Problem Description}

A special setup was used to measure the SPD parameters. It includes a synchronization system, a laser radiation source, a system of beam splitters, a system of variable optical attenuators with controlled output power, an SPD under examination, and an oscilloscope. All components of the system are controlled by software created in the LabVIEW environment.

In our experiments we observe the effect of occurrence the avalanche triggers at the next gate after the photon arrival gate. We analyze the next possible reasons of occurrence of this effect. Our custom InGaAs/InP SPAD (Vereya SPD and Uliana SPD) based SPD was tested to make a decision on what’s the main reason of this shifted avalanche triggers.

\begin{figure}[h]\centering
	\includegraphics[width=1\textwidth]{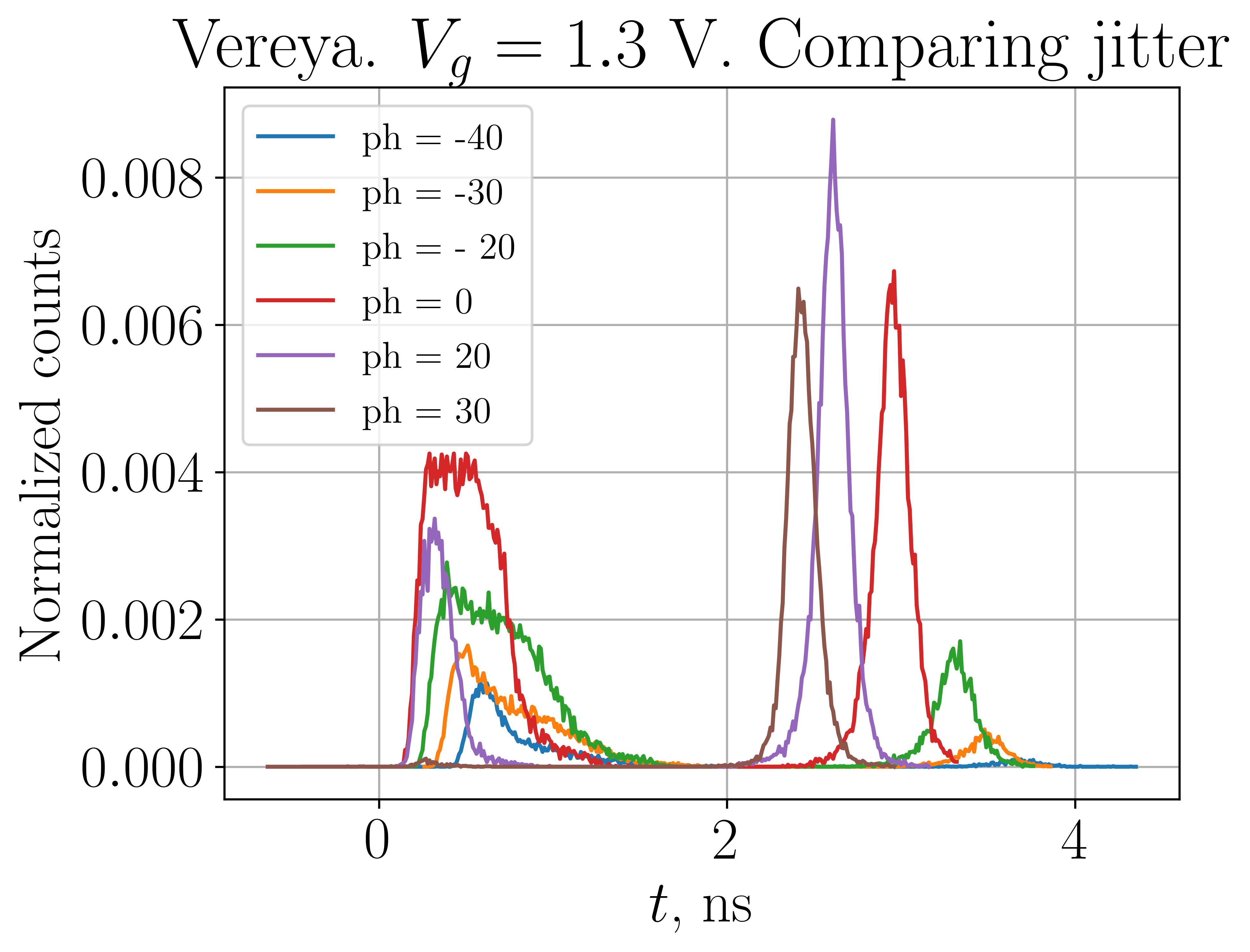}
   \caption{Measured time resolution for Vereya SPD. Gate amplitude $1.3$ V. The histogram is renormalized in according to $PDE$ and is unshifted, but reflected relate to $0$ point}
   \label{fig:vg0_1}
\end{figure}

We tested the Vereya SPD with different gate amplitudes. With low gate amplitude $V_g \approx 1.3$ V (gate code 0) we found that  time resolution has strange form -- it has two separated peaks, with distance between them about $2-3$ ns (see fig. \ref{fig:vg0_1}). We mention that gate period was set to $3.2 ns$, laser pulses with frequency $10$ kHz, and with intensity $1$ ph/pulse. The SPD parameters: $V_b = 72.5$ V, $V_g = 1.3$ V, $V_{br} = 70.9 V$. With this parameters $DCR \approx 1500$ Hz. With the $ph = 0$ we got the maximum count rate $R \approx 5500$ Hz. 

\begin{figure}[h]\centering
	\includegraphics[width=1\textwidth]{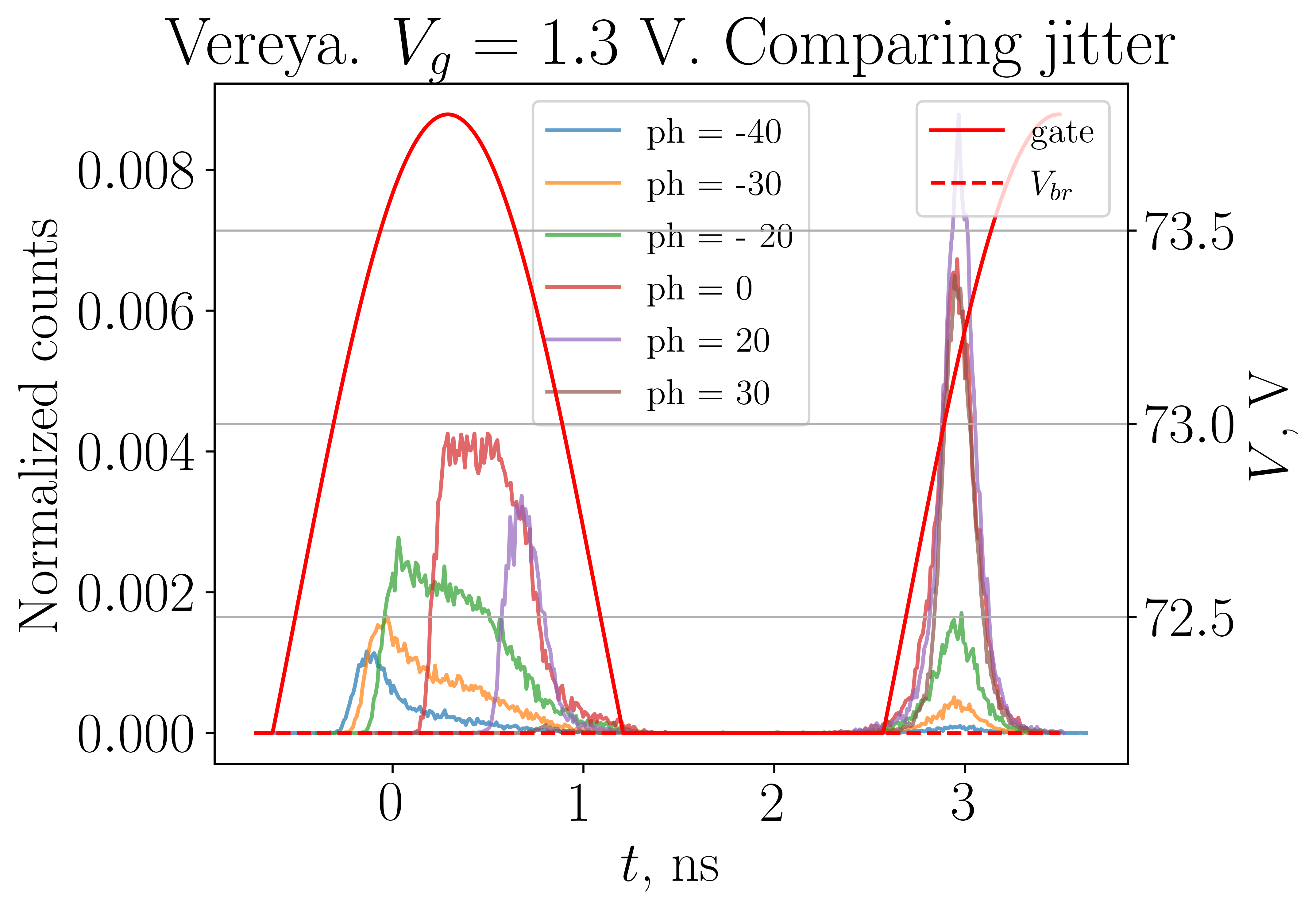}
   \caption{Measured time resolution for Vereya SPD. Gate amplitude $1.3$ V. The $ph = 0$ point denotes the maximum $PDE$. Shifted in according to $ph$.}
   \label{fig:vg0_2}
\end{figure}

The left peaks correspond to the laser pulse detection and determine the simple jitter waveform. The right peaks are undesirable and are due to some strange effect. We can see that with increasing the $ph$, the right peaks shifted to the left. Nature of this effect can be denoted on the figure \ref{fig:vg0_2}.

On this figure we take into account the shifting due to $ph$, and also we plot the gate waveform. We positioned the gate to get the $ph = 0$ near the maximum of the gate voltage, that corresponding to maximum $PDE$.

As we can see on this figure, second peaks persist around one value and overlap each other. We assume, that this effect is determined by some processes at the beginning of the next gate.  

Also, we can see, that for low $ph$ (laser pulse impinge at the beginning of the first gate), the secondary clicks at the second gate are low. But with increasing the $ph$ (shifting the laser pulse to the end of the gate), the secondary peak grows up, but first peak decreases. 

\begin{figure}[h]\centering
	\includegraphics[width=1\textwidth]{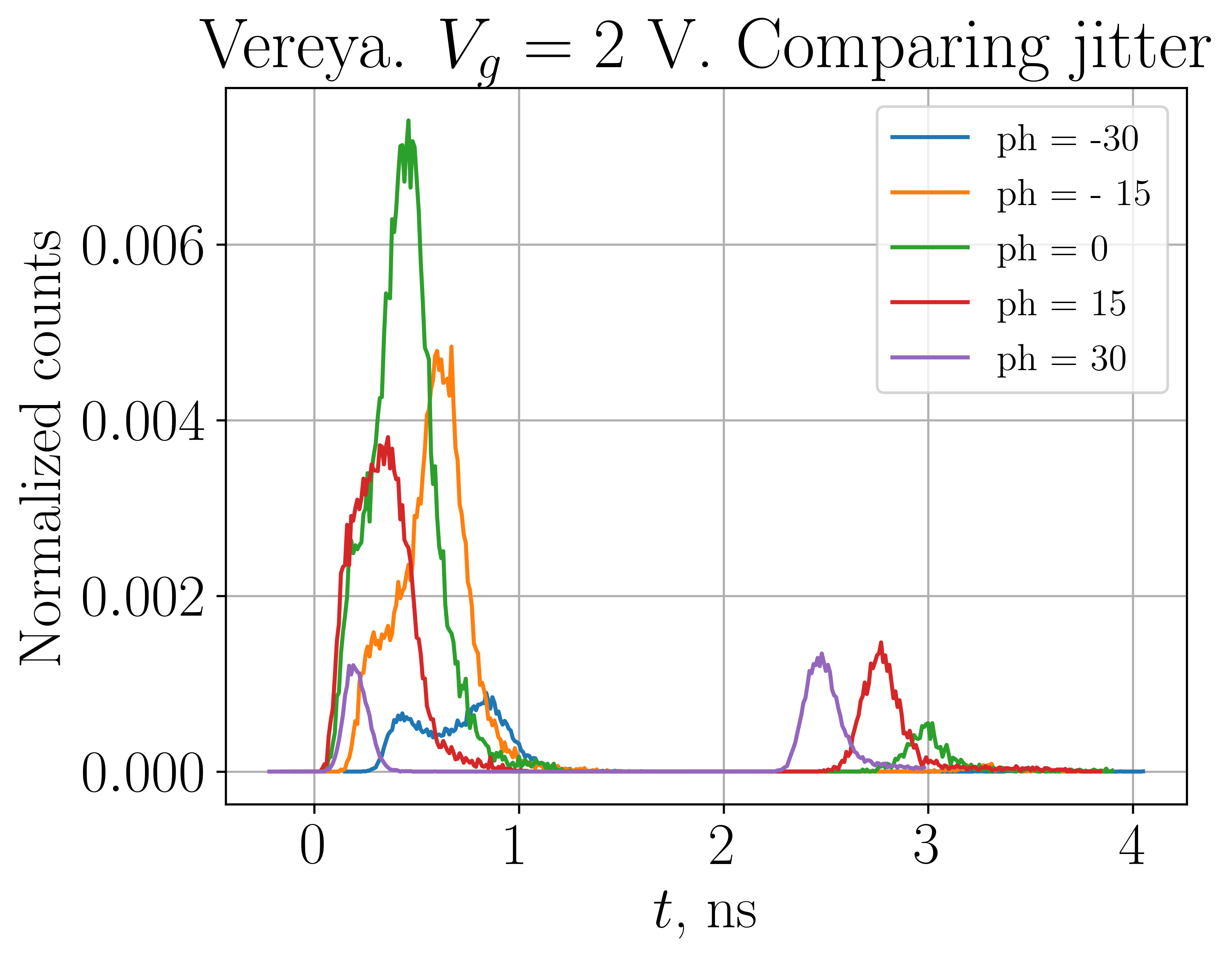}
   \caption{Measured time resolution for Vereya SPD. Gate amplitude $2$ V. The histogram is renormalized in according to $PDE$ and is unshifted, but reflected relate to $0$ point}
   \label{fig:vg99_1}
\end{figure}

Now we think, that this effect can owing to the next reasons:
\begin{itemize}
   \item Afterpulsing effect: the initial (here we mean that this is the charge from first photon or from the low-amplitude and unregistered avalanche) charge trapped at the first gate, and release at the second gate. The approach to verify or folsify is to perform measurements with different temperatures. If histograms didn't change it's form, it means that this effect is not depend on the temperature and is not due to simple trapping. 
   \item Charge persistance effect: The initial charge is trapped at the potential well at the absorption/grading regions heterointerface. This effect should be strongly depend on the gate amplitude and it's bias and virtually not depend on the temperature. If we perform measurements with different maximum $PDE$ and different gate amplitudes $V_g$, and our measurements will not depend, probably this is not charge persistance effect. 
   \item Avalanche delay effect: the initial avalanche is too low, and didn't have time to grow to superior the comparator's level. In this case, there is a lot of free carriers at the structure, that will dissolve, and will not trigger the avalanche due to SPAD stays at the off state. But this free charges (or detrapped charges) will trigger the avalanche at the next gate. We can grow comparator's threshold level $V_{th}$ to increase the probability of comparator triggering by low amplitude avalanches. In this case we will see the lowering the second peak height.  
\end{itemize}

\begin{figure}[h]\centering
	\includegraphics[width=1\textwidth]{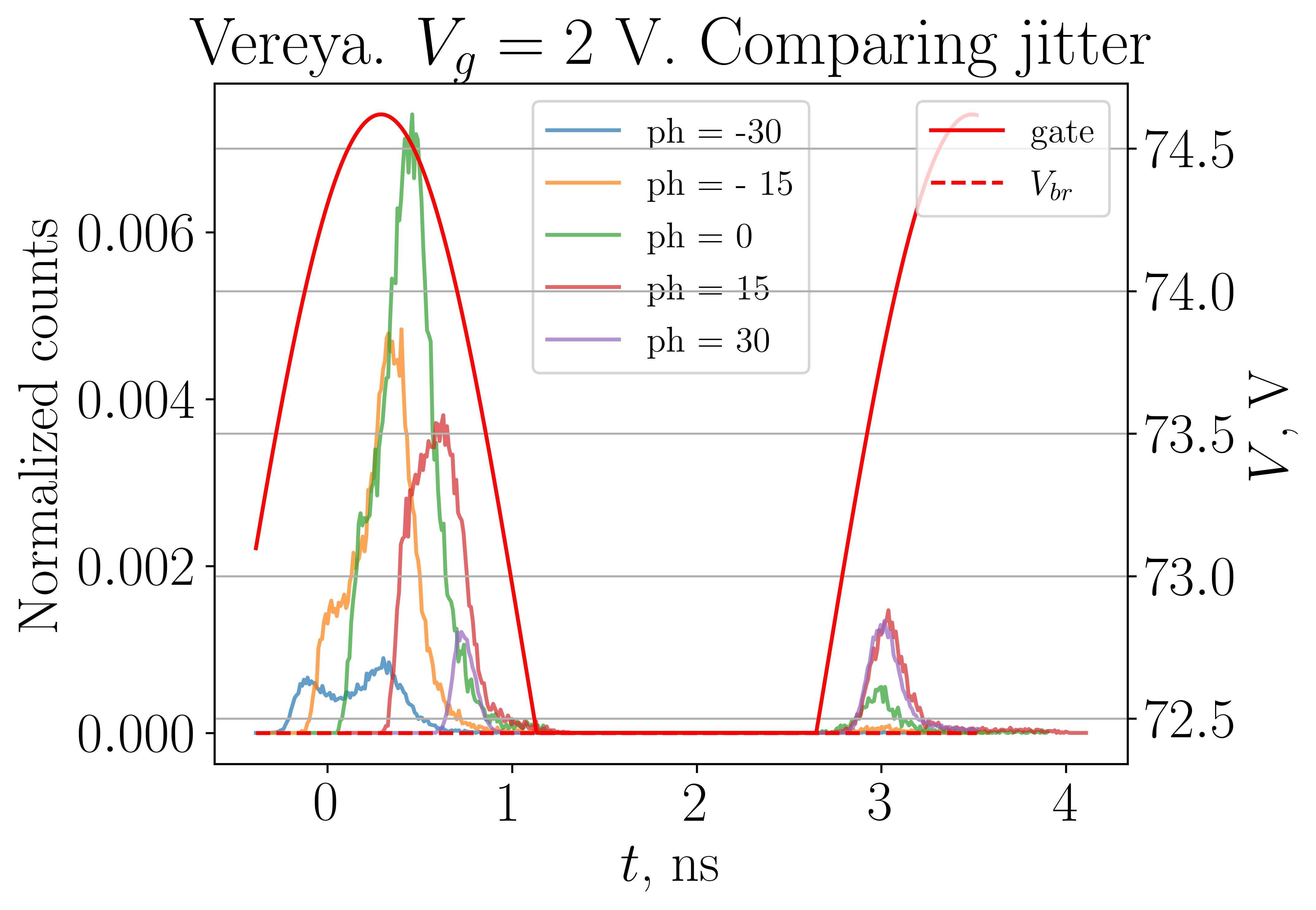}
   \caption{Measured time resolution for Vereya SPD. Gate amplitude $2$ V. The $ph = 0$ point denotes the maximum $PDE$. Shifted in according to $ph$.}
   \label{fig:vg99_2}
\end{figure}

\newpage
\section{Temperature dependence}
To show the temperature dependence of this effect, we take the Uliana SPD, and perform measurements with three different temperatures: $T = -35,\ -45,\ -55 \ {}^\circ$ C. We set the gate code to 0, that corresponds to the peak-to-peak amplitude $\approx 6.5$ V ($V_g = 3.25$ V). We kept the $PDE \approx 23 \ \%$ for all this measurements. But $DCR$ was sufficiently differs for this three temperatures due to different influence of the charge thermal generation effect. $DCR_{-35} = 1500$ Hz, $DCR_{-45} = 850$ Hz, $DCR_{-55} = 300$ Hz.

\begin{figure}[h]\centering
	\includegraphics[width=1\textwidth]{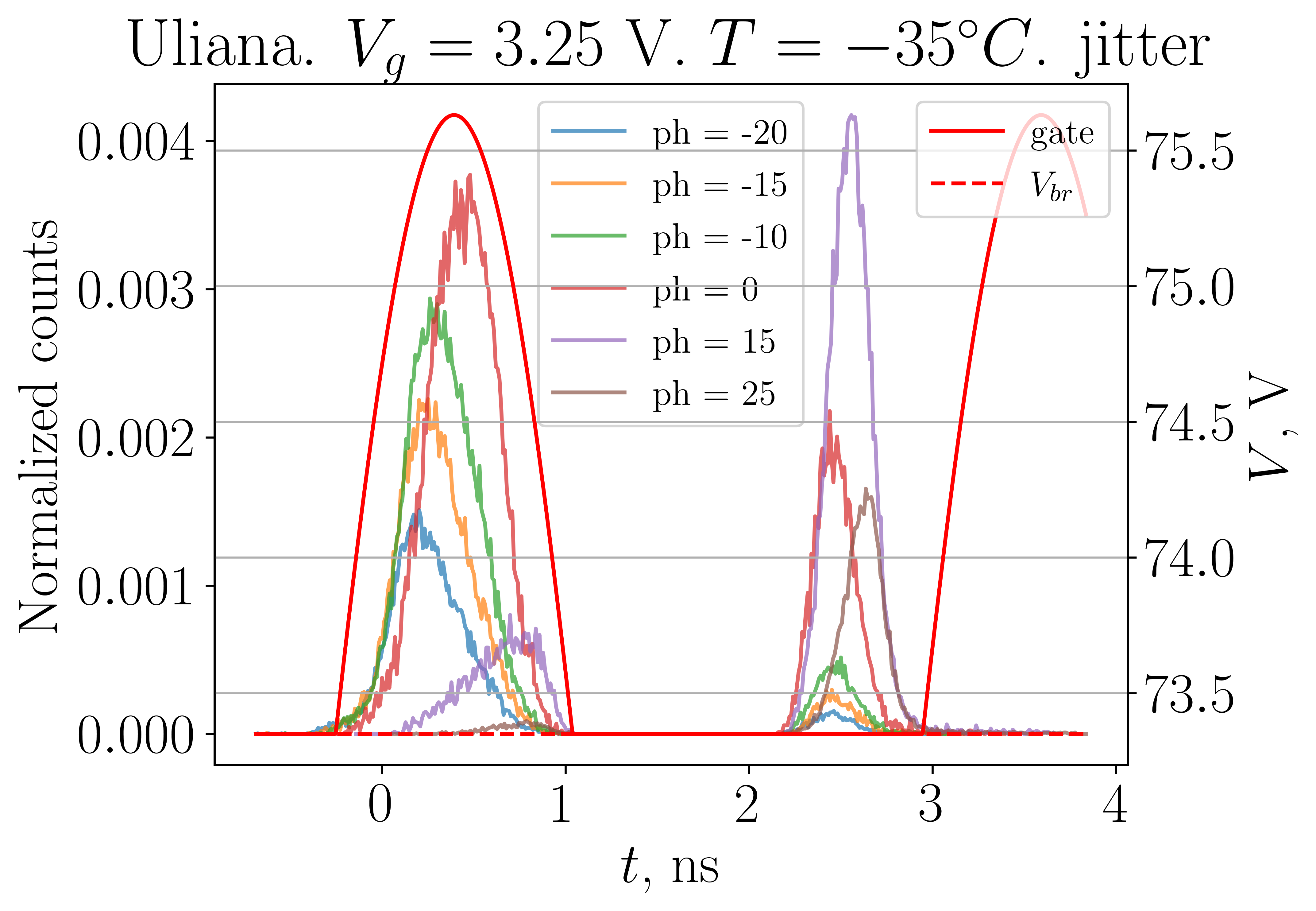}
   \caption{Measured time resolution for Uliana SPD. Gate amplitude $3.25$ V. $T = -35^\circ$ C. The $ph = 0$ point denotes the maximum $PDE$. Shifted in according to $ph$.}
   \label{fig:Uliana_gate0_35_2}
\end{figure}

\begin{figure}[h]\centering
	\includegraphics[width=1\textwidth]{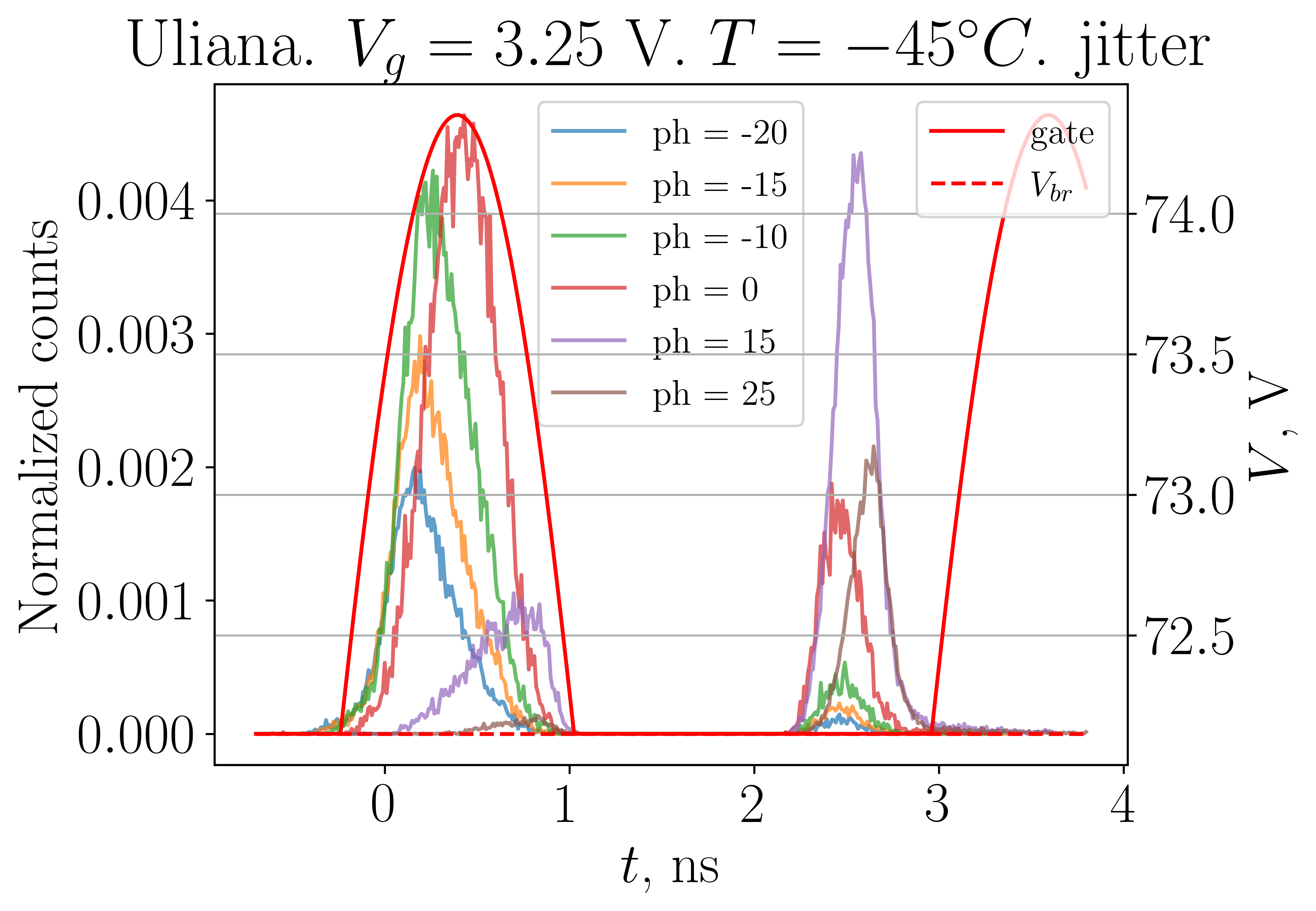}
   \caption{Measured time resolution for Uliana SPD. Gate amplitude $3.25$ V. $T = -45^\circ$ C. The $ph = 0$ point denotes the maximum $PDE$. Shifted in according to $ph$.}
   \label{fig:Uliana_gate0_45_2}
\end{figure}

\begin{figure}[h]\centering
	\includegraphics[width=1\textwidth]{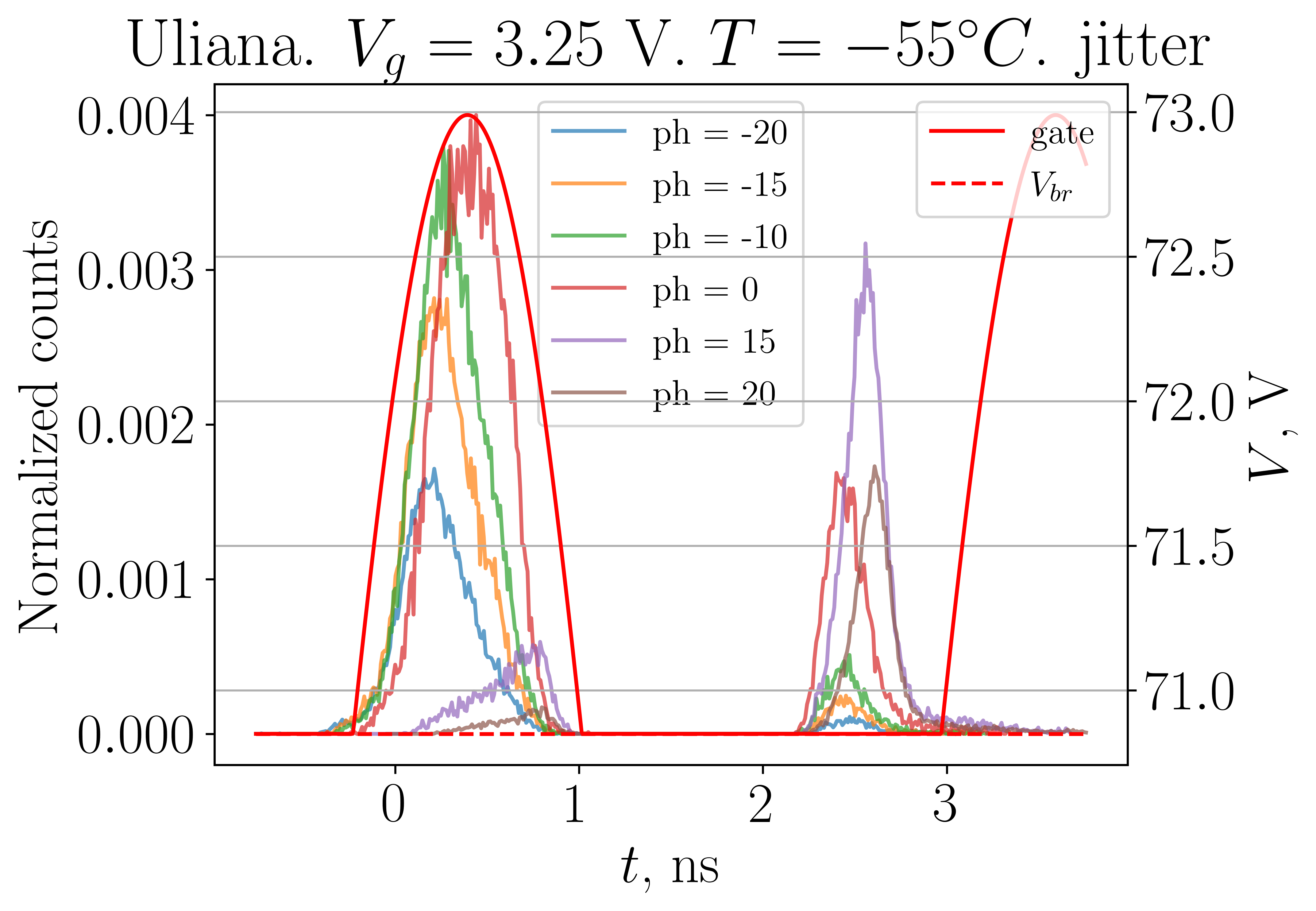}
   \caption{Measured time resolution for Uliana SPD. Gate amplitude $3.25$ V. $T = -55^\circ$ C. The $ph = 0$ point denotes the maximum $PDE$. Shifted in according to $ph$.}
   \label{fig:Uliana_gate0_55_2}
\end{figure}

On the figures \ref{fig:Uliana_gate0_35_2}, \ref{fig:Uliana_gate0_45_2} and \ref{fig:Uliana_gate0_55_2} we presented this time resolution functions. The seconds peaks position is at the off-state of the detector (lefter than gate). It can be due to big uncertainties at the gate amplitude measurements. If we set $V_g \approx 2$ V instead of $3.25$ V, second peaks are in a good position.

\newpage

\newpage

As we can see on this pictures, there is not temperature dependence. All three figures has very similar histogram forms and relative peaks height and positions. So, we can conclude, that this effect is not due to afterpulsing effect. 

\newpage

\section{Gate amplitude dependence}
\begin{figure}[h]\centering
	\includegraphics[width=1\textwidth]{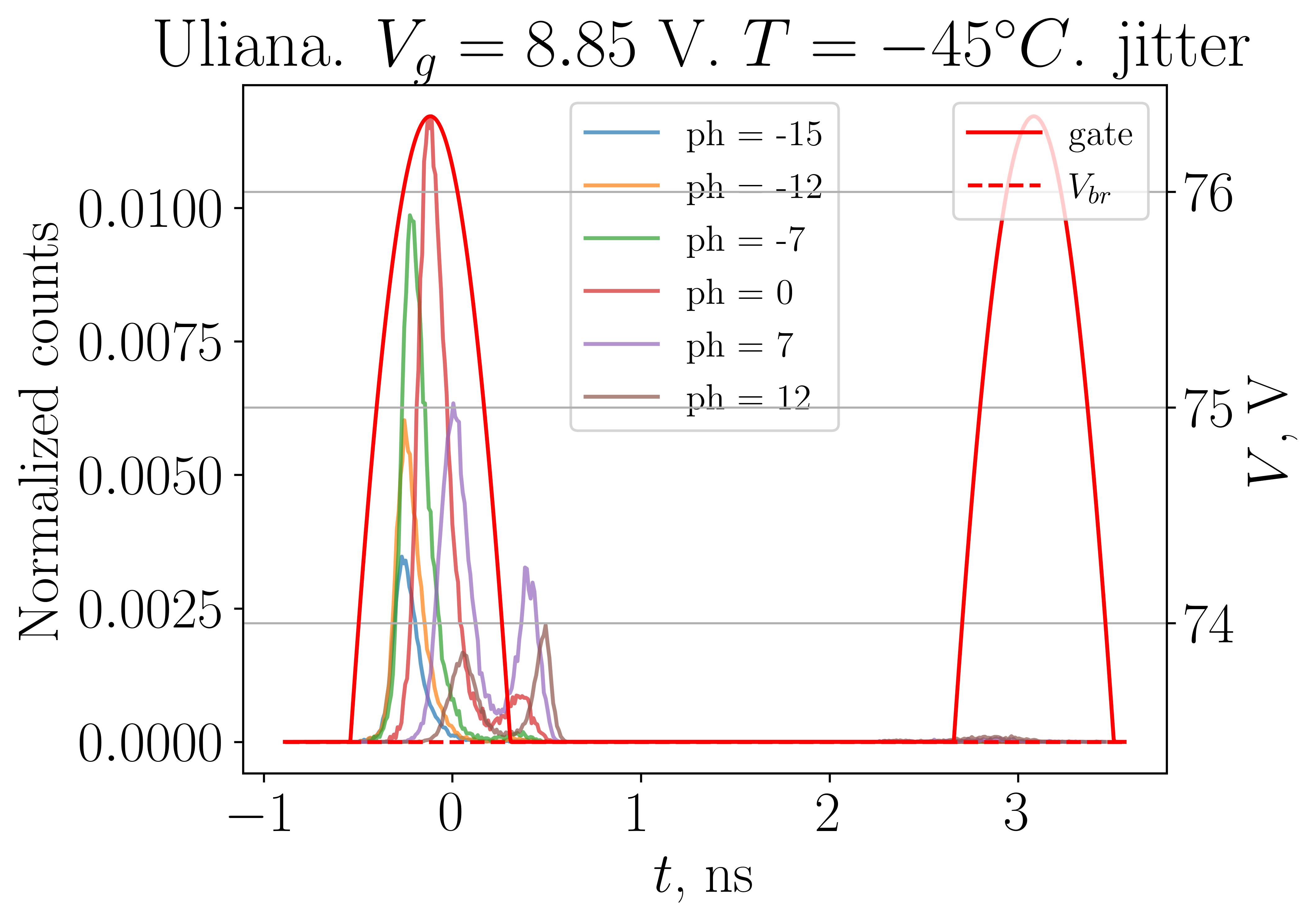}
   \caption{Measured time resolution for Uliana SPD. Gate amplitude $3.25$ V. $T = -55^\circ$ C. The $ph = 0$ point denotes the maximum $PDE$. Shifted in according to $ph$.}
   \label{fig:Uliana_gate3ff_45_2}
\end{figure}

On the pictures \ref{fig:vg0_2} and \ref{fig:vg99_2} we saw, that increasing the gate amplitude leads to the sufficiently decrease of the second peaks height. We perform the similar measurements for Uliana SPD also at the gate code 3ff ($V_g \approx 8.85$ V), and with $PDE \approx 23 \ \%$ at the $T \approx 45^\circ$ C. With this setup $DCR \approx 300$ Hz. The measurements results presented at the picture \ref{fig:Uliana_gate3ff_45_2}.

On this figure we can see that second peaks amplitude is quite low, relate to the $V_g = 3.25$ V figures. It means, that this effect strongly depends on the gate amplitude. 

We can conclude, that effect of triggering photon at the  next gate has more manifistation at the low gate voltages. On the countrary, charge persistance effect has more manifistations at the high gate amplitude, as we conclude at paper about gates. 

In this case, we have only one hypothesis, that we can verify: avalanche delay effect.

\newpage

\section{Bias voltage dependence}
\begin{figure}[h]\centering
	\includegraphics[width=1\textwidth]{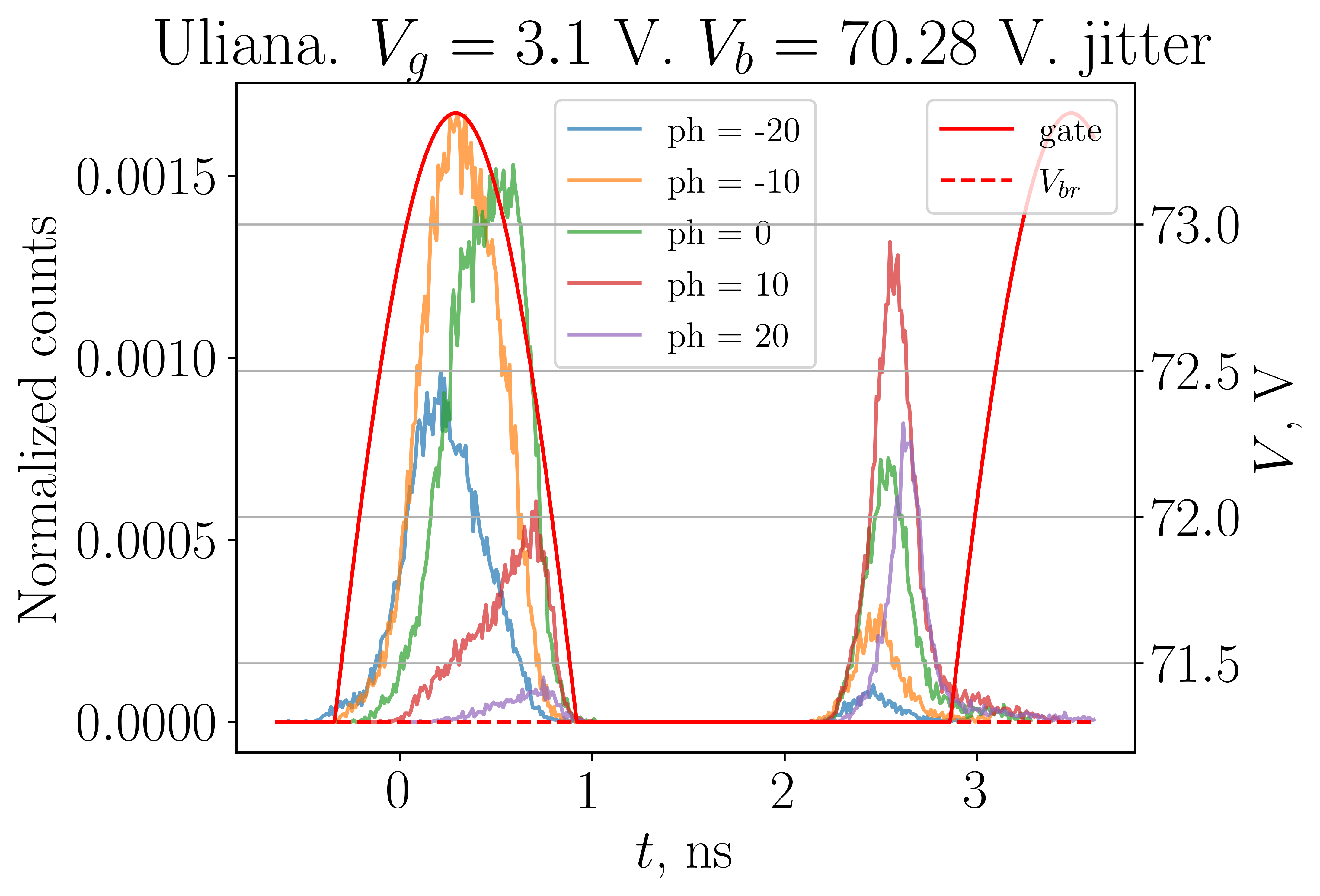}
   \caption{Measured time resolution for Uliana SPD. Gate amplitude $3.25$ V. $T = -50^\circ$ C. Bias voltage $V_b = 70.28$ V. The $ph = 0$ point denotes the maximum $PDE$. Shifted in according to $ph$.}
   \label{fig:Uliana_gate0_vb70.28_2}
\end{figure}

\begin{figure}[h]\centering
	\includegraphics[width=1\textwidth]{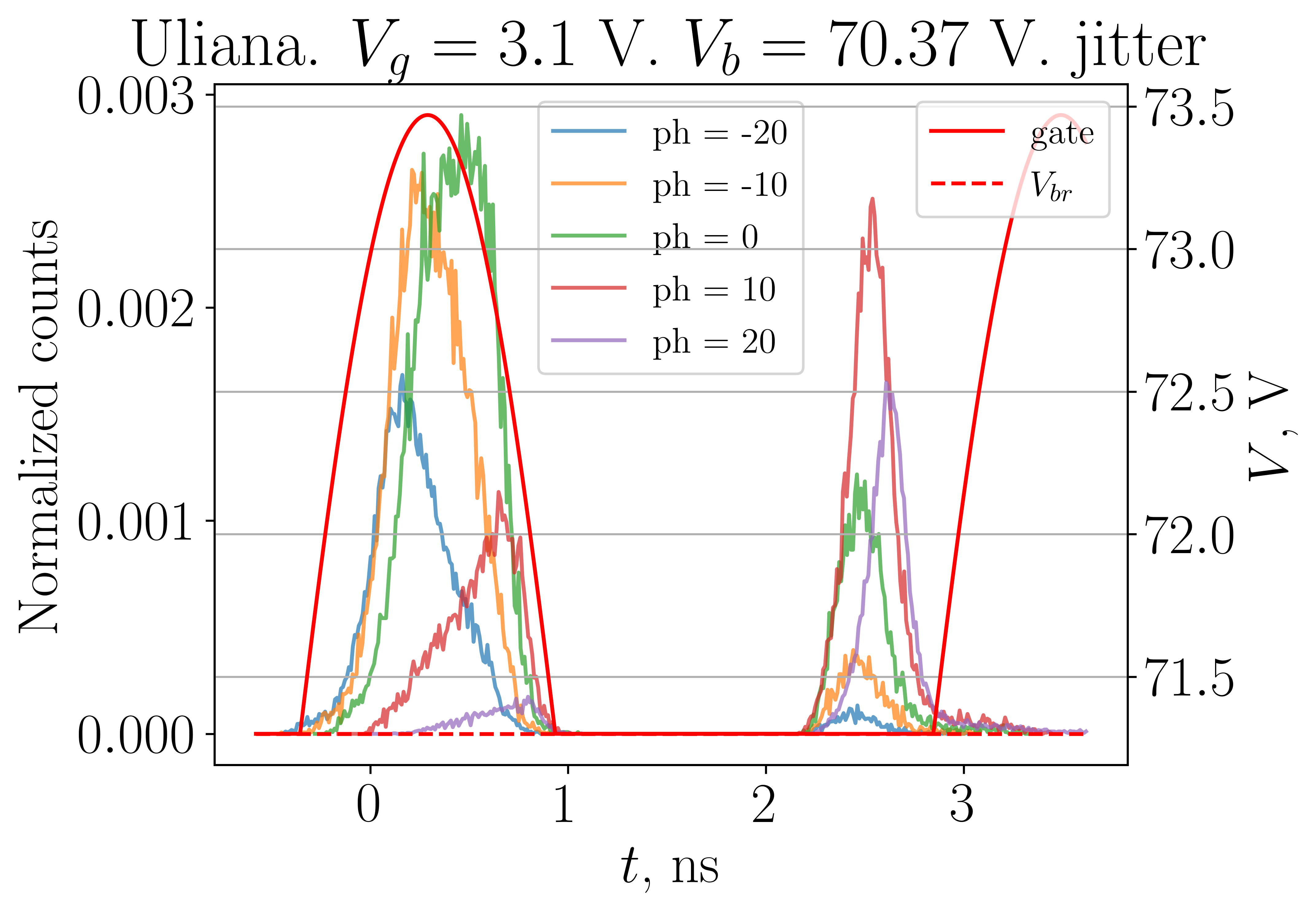}
   \caption{Measured time resolution for Uliana SPD. Gate amplitude $3.25$ V. $T = -50^\circ$ C. Bias voltage $V_b = 70.37$ V. The $ph = 0$ point denotes the maximum $PDE$. Shifted in according to $ph$.}
   \label{fig:Uliana_gate0_vb70.37_2}
\end{figure}

\begin{figure}[h]\centering
	\includegraphics[width=1\textwidth]{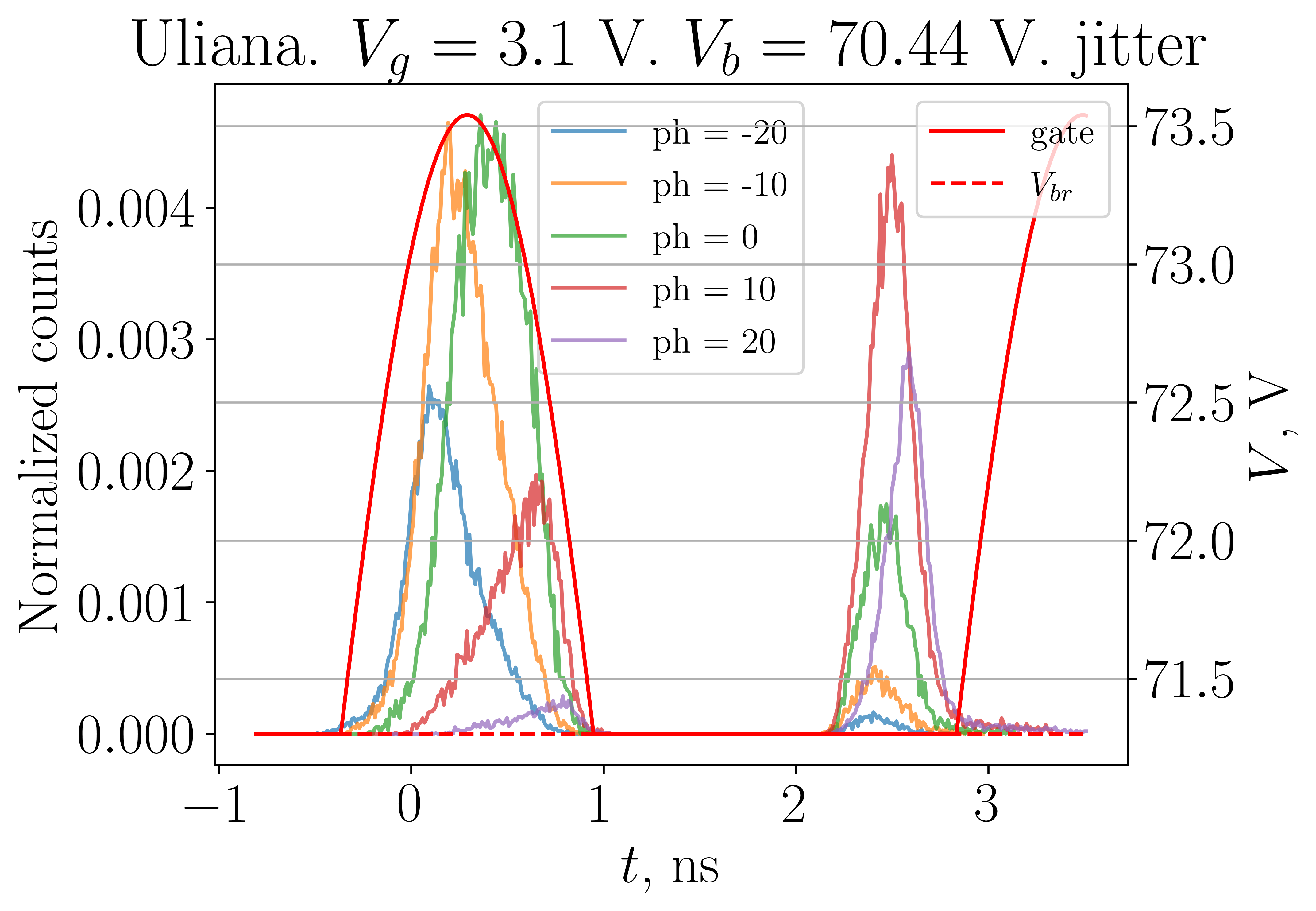}
   \caption{Measured time resolution for Uliana SPD. Gate amplitude $3.25$ V. $T = -50^\circ$ C. Bias voltage $V_b = 70.44$ V. The $ph = 0$ point denotes the maximum $PDE$. Shifted in according to $ph$.}
   \label{fig:Uliana_gate0_vb70.44_2}
\end{figure}

Bias voltage determine the maximum $PDE$ on the SPD. On the Uliana SPD we performed measurements with maximum $PDE \approx 23 \ \%$. Now, we will perform measurements with maximum $PDE \approx 27 \ \%$, $\approx 17.5 \ \%$ and $PDE \approx 9.5 \ \%$ at the temperature $T = -50^\circ$ C and gate code 0 ($V_g = 3.25$ V). The $DCR$ was obtained $600$ Hz, $350$ Hz and $240$ Hz correspondingly. This measurements presented at the figures  \ref{fig:Uliana_gate0_vb70.28_2}, \ref{fig:Uliana_gate0_vb70.37_2}, \ref{fig:Uliana_gate0_vb70.44_2}.

\newpage

\newpage

On this figures we can see, that increasing the bias voltage $V_b$ don't changing the time resolution waveform: first and second peaks has the aprroximately constant relatively height and form. 

\newpage

\section{Comparator's threshold level dependence}
\begin{figure}[h]\centering
	\includegraphics[width=1\textwidth]{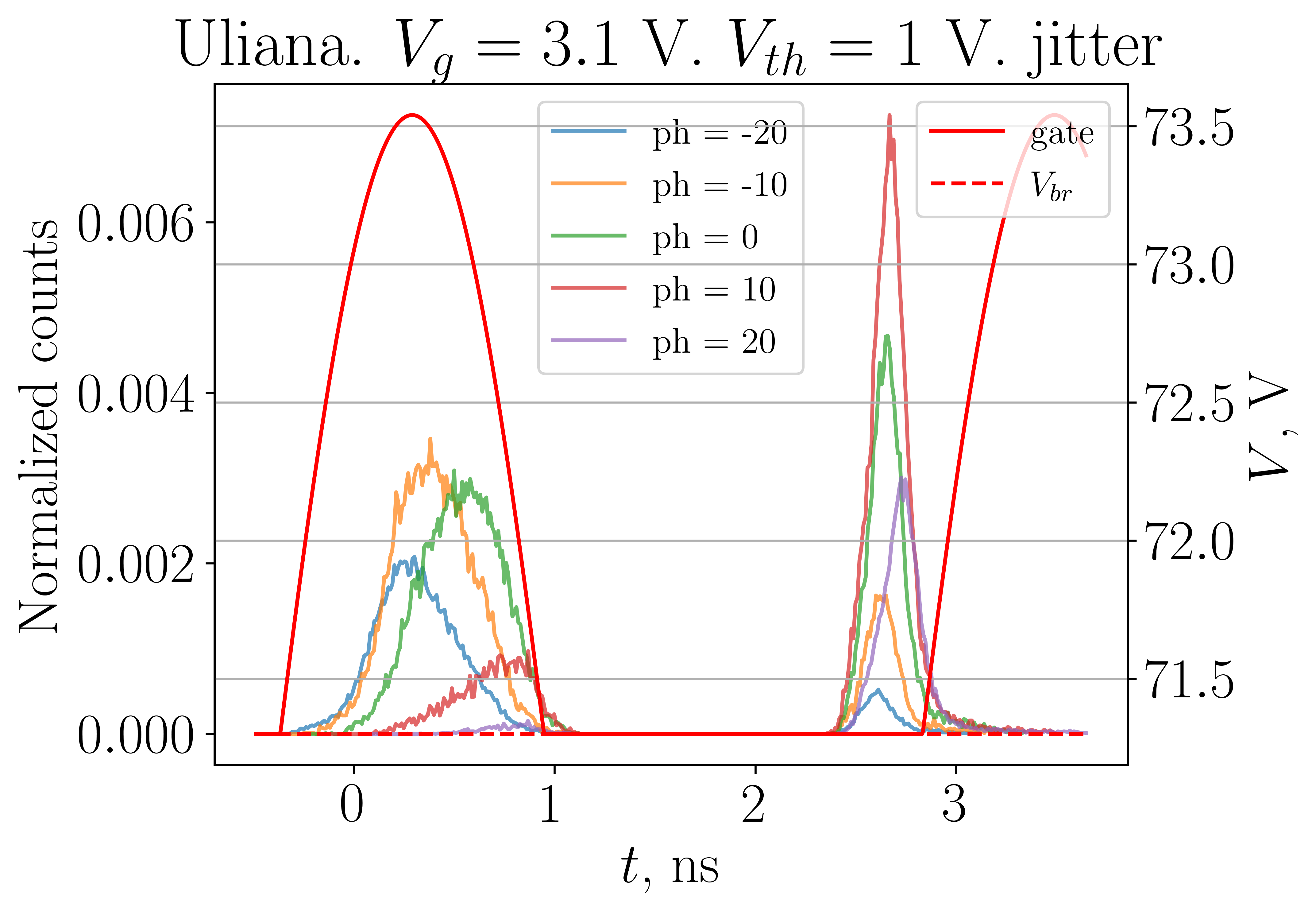}
   \caption{Measured time resolution for Uliana SPD. Gate amplitude $3.25$ V. $T = -50^\circ$ C. Bias voltage $V_b = 70.44$ V. Comparator's threshold voltage $V_{th} = 1$ V. The $ph = 0$ point denotes the maximum $PDE$. Shifted in according to $ph$.}
   \label{fig:Uliana_gate0_vth1_2}
\end{figure}

\begin{figure}[h]\centering
	\includegraphics[width=1\textwidth]{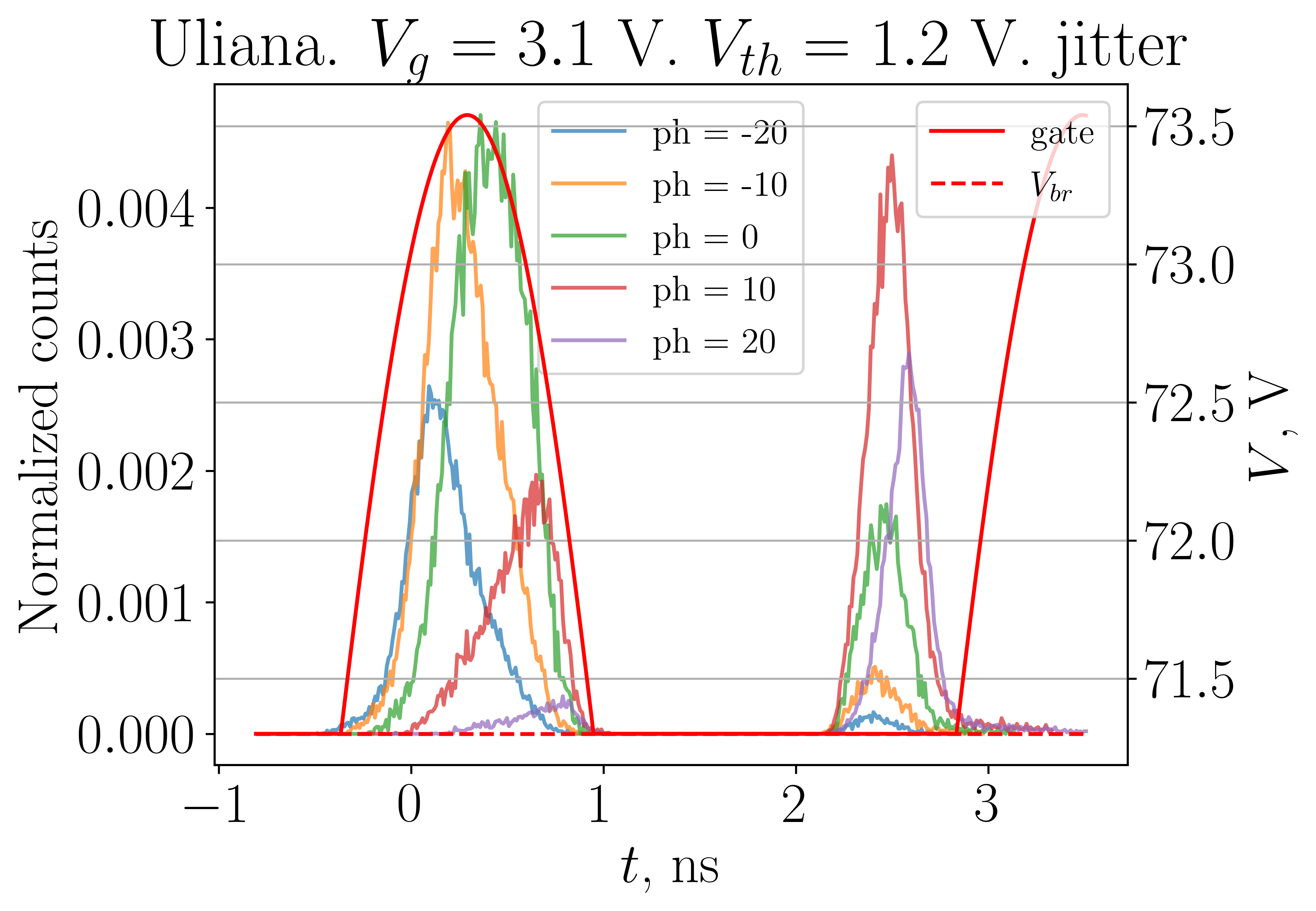}
   \caption{Measured time resolution for Uliana SPD. Gate amplitude $3.25$ V. $T = -50^\circ$ C. Bias voltage $V_b = 70.44$ V. Comparator's threshold voltage $V_{th} = 1.2$ V. The $ph = 0$ point denotes the maximum $PDE$. Shifted in according to $ph$.}
   \label{fig:Uliana_gate0_vth1.2_2}
\end{figure}

\begin{figure}[h]\centering
	\includegraphics[width=1\textwidth]{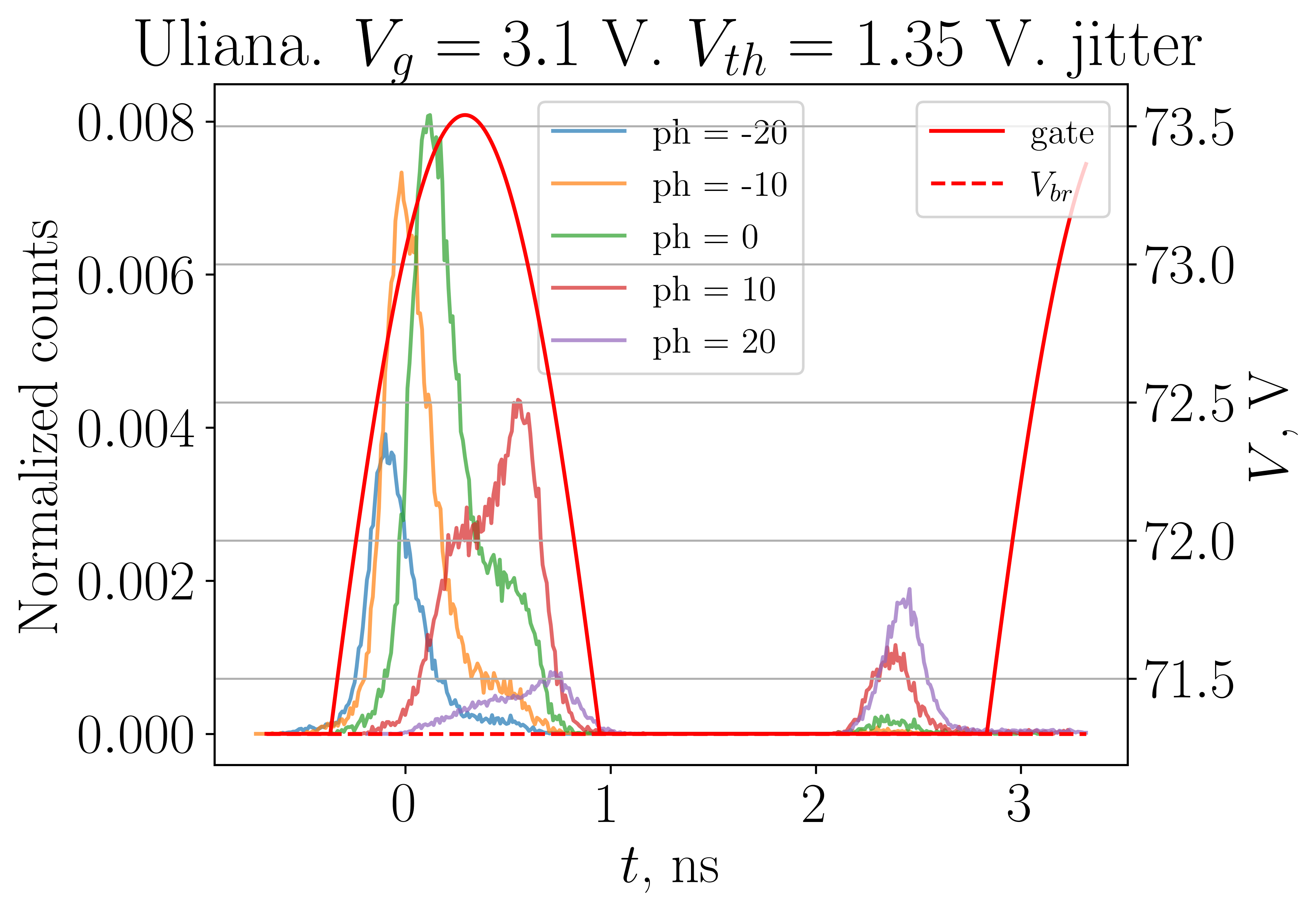}
   \caption{Measured time resolution for Uliana SPD. Gate amplitude $3.25$ V. $T = -50^\circ$ C. Bias voltage $V_b = 70.44$ V. Comparator's threshold voltage $V_{th} = 1.35$ V. The $ph = 0$ point denotes the maximum $PDE$. Shifted in according to $ph$.}
   \label{fig:Uliana_gate0_vth1.35_2}
\end{figure}

Now we perform measurements with different comparator's threshold levels. We make measurements on the Uliana SPD with gate code 0 ($V_g = 3.25$ V) and temperature $T = -50$. Bias voltage was set to $V_b = 70.44$ V, that corresponds to $PDE \approx 27 \ \%$ The shifted voltage with avalanche signal was set to $2$ V. We tested the comparator's threshold voltage: $V_{th} = 1, \ 1.2, \ 1.35$ V. Then more this voltage is, than more probability, that small avalanche will trigger the comparator and consequently accounted. We present this measurements results on the figures \ref{fig:Uliana_gate0_vth1_2}, \ref{fig:Uliana_gate0_vth1.2_2}, \ref{fig:Uliana_gate0_vth1.35_2}.

\newpage

\newpage

On these figures we can see strong dependence of second peak height on the comparathor's threshold voltage: with increasing this voltage, the second's peak height lowers. Now, we can conclude, that this effect is due to avalanche delay effect, described above. The sense of this effect is next: the avalanche at the end of the gate has no time to growing enough to be registered, and continue growing at the next gate, if has not been quenched by high gate voltage. Methods of dealing with this effect  can be the next: increasing the gate voltage and increasing the comparator's threshold level. 

\newpage
\section{Conclusion}

We can conclude, that this effect is due to avalanche delay effect. The sense of this effect is the next: the avalanche at the end of the gate has no time to growing enough to be registered, and continue growing at the next gate, if has not been quenched by high gate voltage.

This is why we obtain lowering of the second peak with increasing the gate amplitude: quenching speed was higher, because applied voltage at off state was lower.

Methods of dealing with the effect of delayed avalanche can be the next: increasing the gate voltage and increasing the comparator’s threshold level.

Further research will be carried out to estimate the lifetime of charge carriers in deep centers in functional regions to substantiate the detected delayed avalanche effect.

\newpage
\bibliographystyle{utphys}
\bibliography{bibliography}

\end{document}